\documentclass[twocolumn,preprintnumbers,amsmath,amssymb,prr]{revtex4}
\usepackage{graphicx}
\usepackage{dcolumn}
\usepackage{bm}
\usepackage{color}
\usepackage{ulem}

\begin{document}

\title{Non-Fermi liquid transport in the vicinity of nematic quantum critical point of FeSe$_{1-x}$S$_x$ superconductor}

\author{W.\,K. Huang$^{1,*}$}
\author{S. Hosoi$^{2,*}\footnote{Present address: Graduate School of Engineering Science, Osaka University\\ 
				\hspace*{-8pt}$^\ast$ The auhors contributed equally to this work.}$}%
\author{M. \v{C}ulo$^{3}$}
\author{S. Kasahara$^1$}
\email{kasa@scphys.kyoto-u.ac.jp}
\author{Y. Sato$^1$}
\author{K. Matsuura$^2$}
\author{Y. Mizukami$^2$}
\author{M. Berben$^3$}
\author{N.\,E. Hussey$^{3,5}$}
\author{H. Kontani$^4$}
\author{T. Shibauchi$^2$}
\author{Y. Matsuda$^1$}
\email{matsuda@scphys.kyoto-u.ac.jp}

\affiliation{$^1$Department of Physics, Kyoto University, Sakyo-ku, Kyoto 606-8502, Japan}
\affiliation{$^2$Department of Advanced Materials Science, University of Tokyo, Kashiwa, Chiba 277-8561, Japan}
\affiliation{$^3$High Field Magnet Laboratory (HFML-EMFL) and Institute for Molecules and Materials, Radboud University, 6525 ED Nijmegen, The Netherlands}
\affiliation{$^4$Department of Physics, Nagoya University, Nagoya 464-8602, Japan}
\affiliation{$^5$H.H. Wills Physics Laboratory, University of Bristol, Tyndall Avenue, BS8 1TL, United Kingdom}

\date{\today}

\begin{abstract}
{Non-Fermi liquids are strange metals whose physical properties deviate qualitatively from those of conventional metals due to strong quantum fluctuations. In this paper, we report transport measurements on the FeSe$_{1-x}$S$_x$ superconductor, which has a quantum critical point of a nematic order without accompanying antiferromagnetism. We find that in addition to a linear-in-temperature resistivity $\rho_{xx}\propto T$, which is close to the Planckian limit, the Hall angle varies as $\cot \theta_{\rm H} \propto T^2$ and the low-field magnetoresistance is well scaled as $\Delta\rho_{xx}/\rho_{xx}\propto \tan^2 \theta_{\rm H}$ in the vicinity of the nematic quantum critical point.  This set of anomalous charge transport properties shows striking resemblance with those reported in cuprate, iron-pnictide and heavy fermion superconductors, demonstrating that the critical fluctuations of a nematic order with ${\bm q} \approx 0$ can also lead to a breakdown of the Fermi liquid description.}
\end{abstract}
	
\maketitle

\section{Introduction}
One of the most prominent features in  many  strongly correlated superconductors is that critical antiferromagnetic (AFM) fluctuations emanating from an AFM quantum critical point (QCP)  seriously modify the quasiparticle masses and scattering cross-section of the Fermi liquid, leading to non-Fermi liquid (NFL) behaviors ~\cite{Lohneysen07,Shibauchi14}. In these systems, the highest superconducting transition temperature $T_c$ is often observed near the QCP.  Therefore, understanding the role of critical fluctuations on the transport properties is an outstanding challenge in condensed matter physics. The resistivity, Hall coefficient and magnetoresistance are the most fundamental transport parameters. It has been reported that all these quantities exhibit a striking deviation from the expected Fermi liquid behavior near the AFM QCP. In conventional metals, the resistivity at low temperatures is often dominated by inelastic electron-electron  scattering. The resistivity from this mechanism varies as $\rho_{xx}\propto T^2$, which is taken as an experimental hallmark of a Fermi liquid. One of the most famous anomalies in the transport properties of  the cuprate and iron-pnictide superconductors as well as the archetypal heavy-fermion systems is a resistivity that evolves linearly with temperature $\rho_{xx}\propto T$~\cite{Martin90, Chien91, Harris95, Kasahara10,Shibauchi14,Nakajima07}, and is close to the Planckian limit~\cite{Bruin13,Legros19,Licciardello19a}.  Moreover, both a strongly $T$-dependent Hall coefficient and a violation of the Kohler's rule in magnetoresistance have been observed in iron-pnictides, such as BaFe$_2$(As$_{1-x}$P$_x$)$_2$~\cite{Kasahara10}, the two-dimensional (2D) heavy fermion superconductors Ce$M$In$_5$ ($M = $ Co and Rh)~\cite{Nakajima07} near the AFM QCP~\cite{Shibauchi14} and in cuprates within the optimally doped and underdoped regimes \cite{Chien91,Harris95}, which  have been discussed in terms of NFL properties \cite{Stojkovic97,Kontani99,Kontani08}.

Recently, electronic nematicity that spontaneously breaks the rotational symmetry of the underlying crystal lattice has been a growing issue in strongly correlated electron systems \cite{Fradkin10}, including cuprates \cite{Sato17, Murayama19,Ishida19,Auvray19}, iron pnictides/chalcogenides~\cite{Fernandes14,Chu10,Kasahara12} and heavy fermion compounds~\cite{Ronning17}. Understanding the role of the critical quantum fluctuations of nematic order on the normal and superconducting states is of utmost importance~\cite{Fernandes14,Lederer17}.   The layered compound FeSe$_{1-x}$S$_x$ offers a fascinating opportunity to study this issue~\cite{SHM}.  FeSe is a compensated semimetal, which exhibits a structural transition from tetragonal to orthorhombic  symmetry at $T_s\approx 90$\,K . Below $T_s$, no magnetic order occurs in contrast to the other iron-based compounds~\cite{Baek15,Bohmer15}. The nematicity can be tuned continuously by isoelectronic sulfur substitution. At $x\approx$ 0.17, $T_s$ vanishes and the nematic susceptibility diverges, indicating the presence of a nematic QCP~\cite{Hosoi16}. The AFM fluctuations are also suppressed by the  substitution and no sizable AFM fluctuations are observed near the nematic QCP~\cite{Wiecki17}. 
  
Although the Fermi surface changes smoothly when crossing the nematic QCP in FeSe$_{1-x}$S$_x$~\cite{Coldea19, Hanaguri18},  the nematic QCP has a significant impact on both the  superconducting~\cite{Sato18, Hanaguri18} and normal state properties~\cite{Licciardello19a, Urata16, Bristow19, Licciardello19b}. The superconducting gap structure exhibits a dramatic change at the QCP. The in-plane resistivity exhibits a $T$-linear behavior, $\rho_{xx}\propto T$~\cite{Licciardello19a, Bristow19, Urata16}, at low temperature in the vicinity of the nematic QCP, suggesting  that the critical fluctuations of nematic order  can give rise to deviations from Fermi liquid behavior~\cite{Lederer17,Berg19}. However, the $T$-linear $\rho_{xx}$ alone cannot be simply taken as conclusive evidence for the NFL behavior because it can appear as a result of multiband effects (see Appendix).  Therefore, measurements of the Hall coefficient and magnetoresistance across the nematic QCP of FeSe$_{1-x}$S$_x$ are of pivotal importance in understanding the role of the nematic critical fluctuations on the transport properties~\cite{Nakajima07,Lederer17,Berg19,Paul17,deCarvalho19}.

In this paper, we measured the low-field transport properties of FeSe$_{1-x}$S$_x$ for $0\leq x \leq 0.22$.  In the vicinity of the nematic QCP at $x\approx 0.17$, the linear-in-temperature resistivity is observed below $\sim 40$\,K, as reported in Ref.~\cite{Licciardello19a}. In these same $x$-- and $T$-- regimes, the Hall coefficient and magnetoresistance exhibit characteristic behaviors that bear resemblance to the anomalous transport behaviors reported in iron-pnictide and heavy fermion superconductors in the vicinity of the AFM QCP~\cite{Kasahara12, Shibauchi14, Nakajima07} and in cuprates in the optimally and underdoped regimes~\cite{Bruin13, Legros19, Licciardello19a, Martin90}. These similarities appear to suggest that essentially different types of critical quantum fluctuations give rise to several common deviations from conventional Fermi liquid behavior near the QCP.

\section{Experimental}
High quality single crystals of FeSe$_{1-x}$S$_x$ were grown by the chemical vapour transport method. We determined $x$ values by energy dissipative X-ray analysis. The crystals in the orthorhombic regime ($0\leq x\leq 0.18$) are heavily twinned. Most of the measurements were performed on samples with typical dimensions of $2 \times 1 \times 0.05\,{\rm mm}^3$.  The electrical current is applied along the $ab$ plane.  The Hall effect and transverse magnetoresistance were measured simultaneously.  We obtained the Hall resistivity $\rho_{xy}$ from the transverse resistance by subtracting the positive ({\boldmath $H$} $ \parallel c$) and negative magnetic field ({\boldmath $H$} $ \parallel -c$) data.

\section{Results}
\subsection{Resistivity}
\begin{figure}[t]
	\includegraphics[width=\linewidth]{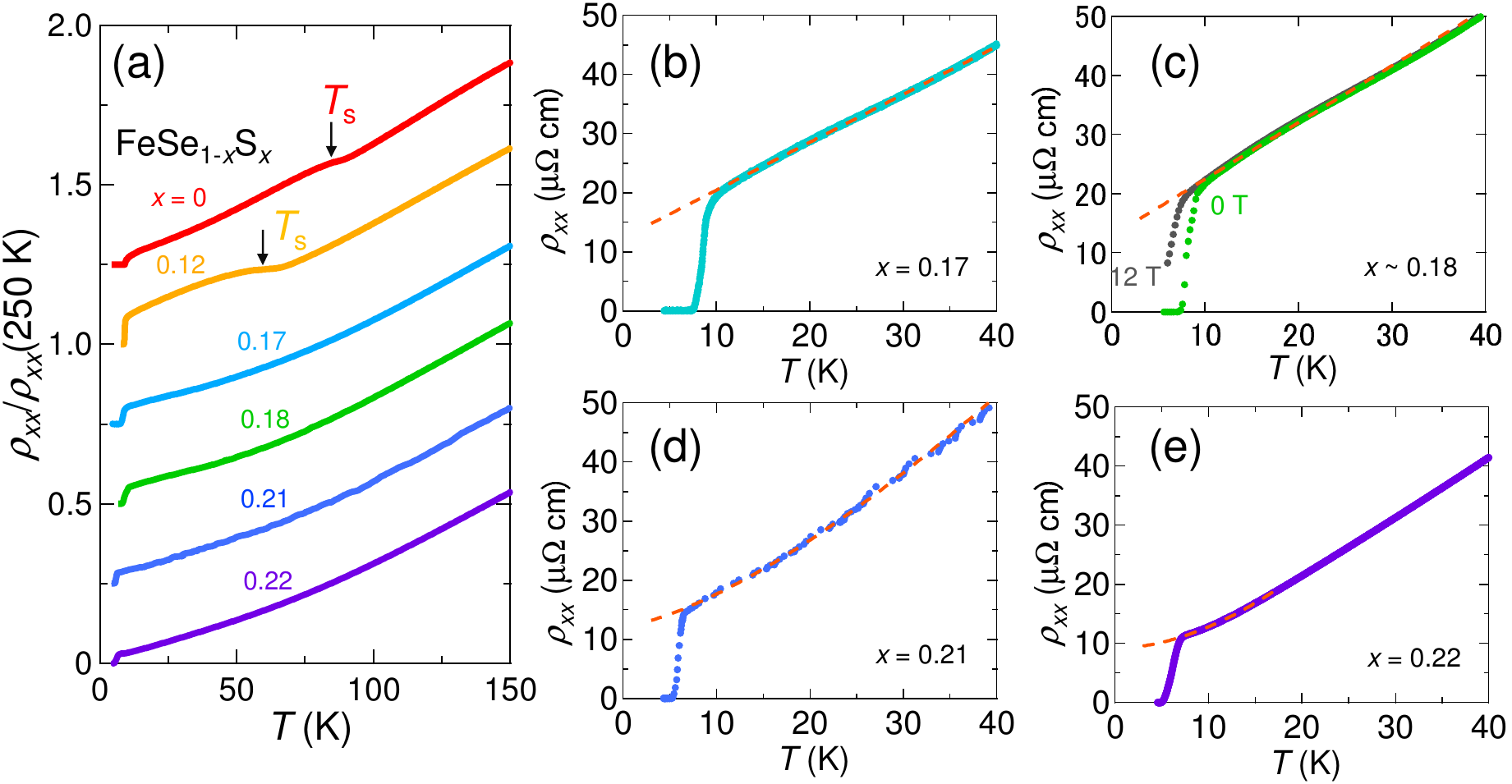}
	\caption{ (a) Temperature dependence of the resistivity $\rho_{xx}$ normalized by $\rho_{xx}$ at 250\,K in zero field of FeSe$_{1-x}$S$_x$ for various  S substitution levels. Arrows indicate the nematic transition temperature $T_s$.  (b) Temperature dependence of $\rho_{xx}$ at low temperatures in zero field for $x=0.17$.  The red dashed line represents a line fitted by $\rho_{xx}\propto T$. (c) The same plot in zero field and at $\mu_0H=12$\,T applied parallel to the $ab$ plane for $x=0.18$.  The red dashed line represents a line fitted by $\rho_{xx}\propto T$.
	(d) The same plot in zero field for $x=0.21$.  The red dashed curve represents a fitting curve $\rho_{xx} \propto T^{1.5}$. (e) The same plot in zero field for $x=0.22$.  The red dashed curve represents a fitting curve $\rho_{xx} \propto T^2$. 
	} 
	\label{resistivity}
\end{figure}

As shown in Fig.\,\ref{resistivity}(a),  $\rho_{xx}$ shows a kink at $T_s$. With increasing $x$, $T_s$ decreases to lower temperature and no kink anomaly is observed at $x\geq0.17$, indicating that the system is in the tetragonal phase. The Fermi surface of FeSe$_{1-x}$S$_x$ consists of hole cylinders around the zone center and compensating electron cylinders around the zone corner~\cite{Kasahara14,Yi19}. In FeSe in the nematic phase, the Fermi energies of both the hole and the electron pockets are extremely small: $\varepsilon_F^h\approx 10$\,meV and $\varepsilon_F^e\sim 5$\,meV, respectively~\cite{Kasahara14}. With sulfur substitution, $\varepsilon_F^h$ increases gradually without exhibiting an abrupt change at the nematic QCP~\cite{Hanaguri18}.  In the tetragonal phase at $x\geq 0.17$, the Fermi surfaces consist of two hole and two electron pockets. Recent angle-resolved-photoemission-spectroscopy (ARPES) measurements report $\varepsilon_F^h\approx14$\,meV for both hole pockets.  Although direct measurements of the electron pockets by ARPES is lacking,   $\varepsilon_F^e$ is speculated to be of order of 5 -- 15\,meV for both electron pockets.  This is because the  electronic specific heat coefficient in the tetragonal phase is close to that in the nematic phase \cite{Sato18}, suggesting no dramatic change of the effective electron mass.  We therefore focus on the transport properties  below $\sim 40$\,K [Figs.\,\ref{resistivity}(b)-(d)], where the temperature is sufficiently below the Fermi energy of each pocket.  To study the $T$-dependence of $\rho_{xx}$ to lower temperatures, a magnetic field is applied parallel to the $ab$ plane to suppress the superconductivity.  In this configuration,  magnetoresistance arising from the orbital motion of the electrons is negligibly small as reported in Ref.~\cite{Licciardello19a}.  The most distinct feature is the linear-in-temperature resistivity typical to NFL behaviour down to low temperatures at $x=0.17$ and 0.18, at or slightly above the nematic QCP. For $x=0.22$, $T$-linear dependence of the resistivity changes to approximately $T^2$-dependence as expected for a Fermi liquid at lower temperatures. Similar behavior has been reported now by several groups~\cite{Urata16, Licciardello19a, Bristow19}.


\begin{figure}[t]
	\includegraphics[width=0.8\linewidth]{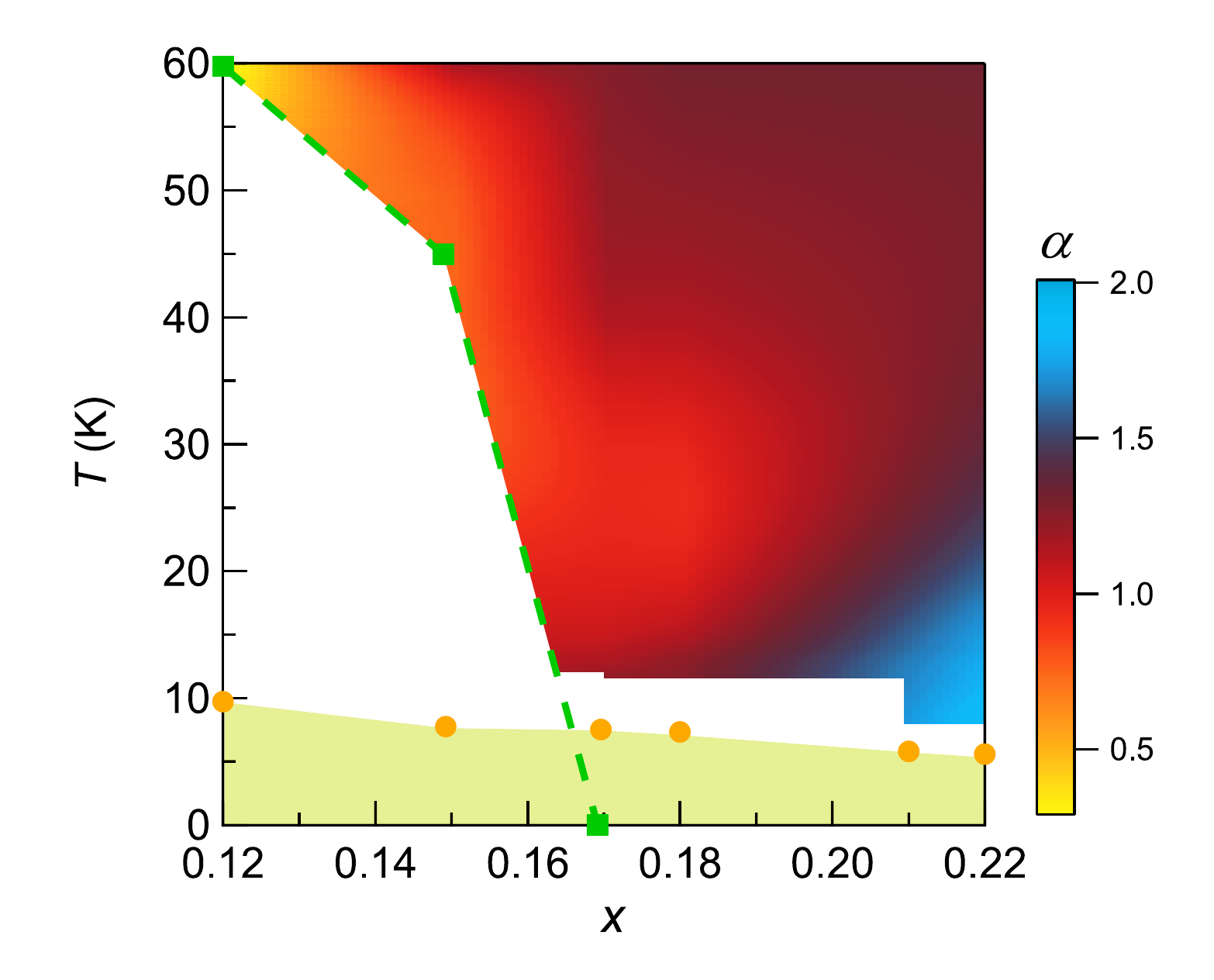}
	\caption{Phase diagram of FeSe$_{1-x}$S$_x$.   	Colors in the tetragonal  state represent evolution of the	exponent $\alpha$
		in resistivity fitted by Eq.\,\ref{alpha}. The green dashed line represents the tetragonal-orthorhombic (nematic) transition temperature $T_s$ determined by the resistivity measurements.  The green square is the nematic QCP.   Orange solid circles represent the superconducting transition temperature. 	
	} 
	\label{Colorplot}
\end{figure}

Figure\,\ref{Colorplot} shows the color plot of the exponent $\alpha$ in the $T$-dependence of the resistivity
\begin{equation}
\rho_{xx}=\rho_{0}+AT^{\alpha}.
\label{alpha}
\end{equation}   
Blue and red show the region in which  Fermi liquid ($\alpha\approx 2$) and  NFL ($\alpha \approx 1$) behaviors are observed, respectively. In the nematic regime below $T_s$, the $T$-dependence of $\rho_{xx}$ exhibits a concave downward curvature, which is likely to be caused by a change in the carrier scattering time $\tau$ associated with orbital ordering at $T_s$.  We do not discuss the $T$-dependence of $\rho_{xx}$  below $T_s$ in detail.  The Fermi liquid regime with $\alpha\approx 2$ is seen in the low temperature regime at large $x$ values~\cite{Licciardello19a}. The phase diagram also includes a funnel of $T$-linear resistivity centered on $x\approx 0.17$, indicating that the critical fluctuations originating from the nematic QCP extend up to a finite temperature.

It has been discussed previously that in cuprates and other correlated systems, the $T$-linear $\rho_{xx}$ appears in the Planckian limit, where $\hbar/\tau=\Lambda k_BT$  with $\Lambda \approx 1$ is fulfilled~\cite{Legros19,Bruin13}.  To examine whether FeSe$_{1-x}$S$_x$  is in the  Planckian limit, we calculate  $A$ per 2D sheet $A^{\square}=A/d$ ($d$ is interlayer distance), which  is given as $\Sigma_i A^{\square}\varepsilon^i_F/k_B = \Lambda (h/2e^2) $. From the $T$-linear $\rho_{xx}$ in the vicinity of nematic QCP, we obtain $A\approx 1.0$\,$ \mu\Omega$cm/K.  As discussed above, the electronic specific heat coefficient in the tetragonal phase does not change dramatically from that in the nematic phase~\cite{Sato18}. Therefore, by simply assuming $\varepsilon_F^{e,h} \sim 10$ -- $15$\,meV for all pockets,  $\Lambda = 0.7$ -- 0.9 is obtained. An earlier analysis of the coefficient of the $T$-linear resistivity using a multi-band Drude model also found $\Lambda \sim 1$  ~\cite{Licciardello19a}, implying that scattering within the quantum critical fan of a nematic QCP is indeed close to the Planckian limit.

\subsection{Hall effect}
\begin{figure}[t]
	\includegraphics[width=\linewidth]{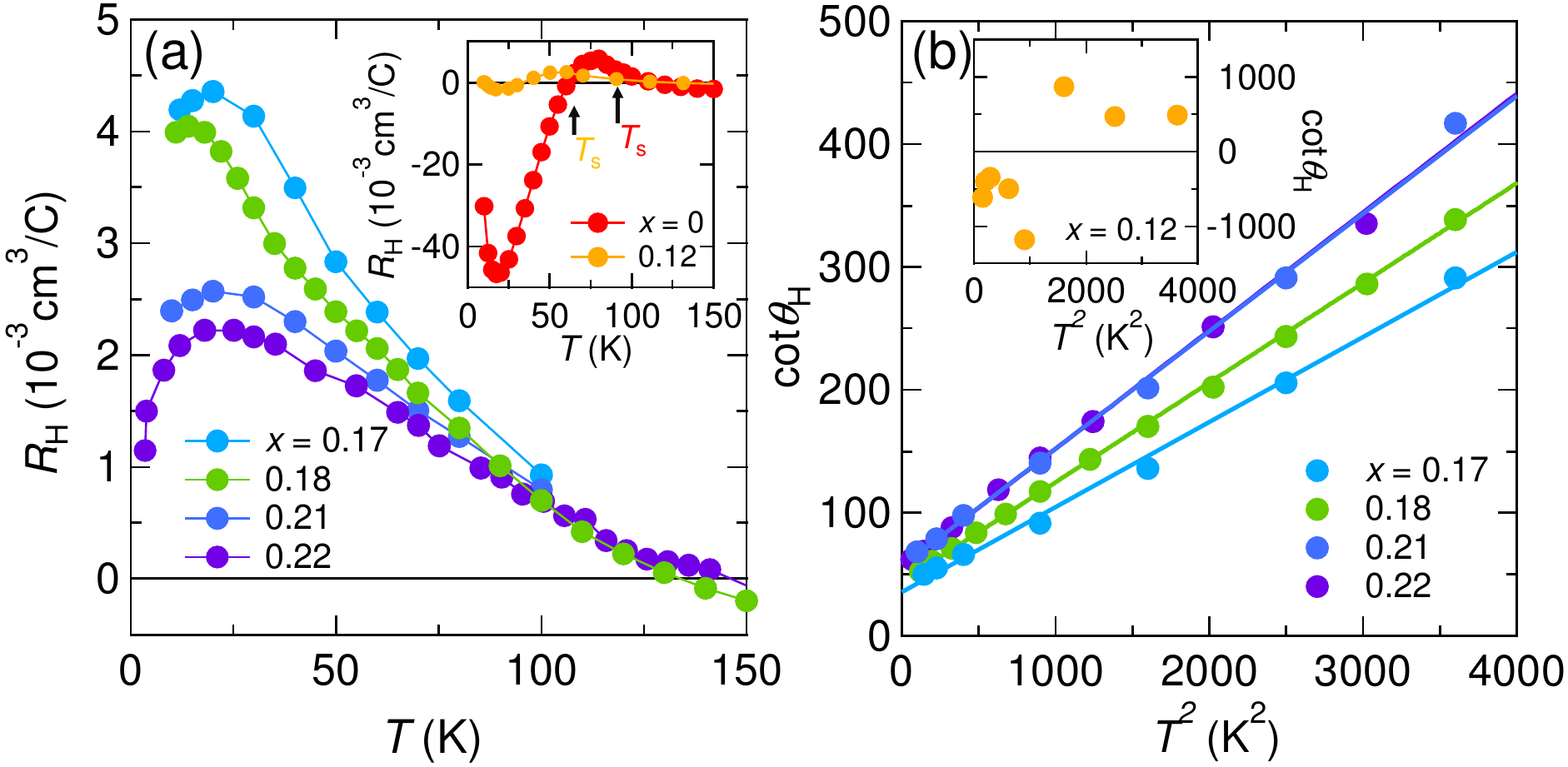}
	\caption{{\bf a}, Temperature dependence of the Hall coefficient $R_{\rm H}$  for $x\geq0.17$.  The inset shows $R_{\rm H}$ for $x<0.17$. Arrows indicate the tetragonal-orthorhombic transition temperature. {\bf b}, Cotangent of the Hall angle plotted as a function of $T^2$ in the tetragonal regime of FeSe$_{1-x}$S$_x$ ($x\geq 0.17$).  The inset shows the same plot in the nematic regime for $x=0.12$.
	} 
	\label{Hall}
\end{figure}

The Hall coefficient $R_{\rm H}$ is determined largely by the Fermi surface topology and carrier density. In the simple case of a Fermi liquid, if there is a single species of charge carrier and $\tau$ is the same at all points on the Fermi surface,  $R_{\rm H} = 1/ne$, where $n$ is the carrier number and $R_{\rm H}$ is independent of temperature. Figure\,\ref{Hall}(a) and its inset depict the $T$-dependence of $R_{\rm H} (\equiv d\rho_{xy}/dH$ at $H \rightarrow 0$) at various $x$. In the nematic regime, $\rho_{xy}$ shows non-linear $H$-dependence, while in the tetragonal regime, $\rho_{xy}$ exhibits nearly perfect $H$-linear dependence.   As shown in the main panel of Fig.\,\ref{Hall}(a), $R_{\rm H}$ is positive in the tetragonal regime at $x \geq 0.17$, indicating the dominant contribution of holes. As the temperature is lowered, $R_{\rm H}$ increases rapidly with increasing slope.    Slightly above $T_c$, $R_{\rm H}$ peaks and decreases. A salient feature is that the enhancement of $R_{\rm H}$ at low temperatures becomes more pronounced upon approaching  $x\approx0.17$, suggesting that the Hall effect is strongly influenced by the  critical nematic fluctuations.

In cuprates~\cite{Chien91} iron-pnictides \cite{Kasahara10} and heavy fermion compounds  \cite{Nakajima07} in the vicinity of AFM QCP, $|R_{\rm H}|$ increases rapidly with decreasing temperature. It has been pointed out that the Hall problem in cuprates and in Ce$M$In$_5$ can be simplified  by analyzing the Hall response in terms of the Hall angle,  $\theta_{\rm H} \equiv \tan^{-1}(\rho_{xy}/\rho_{xx})$~\cite{Chien91, Nakajima07} as,
\begin{equation}
	\cot \theta_{\rm H} = a + bT^2,
	\label{Halleq}
\end{equation} 
where $a$ and $b$ are constants.  This Hall angle behavior,  when combined with the $T$-linear resistivity, is an important hallmark of  the deviations from Fermi liquid transport.

In Fig.\,\ref{Hall}(b), we plot  $\cot \theta_{\rm H}$ of FeSe$_{1-x}$S$_x$ as a function of $T^2$.  For $x < 0.17$, $\cot \theta_{\rm H}$ displays non-monotonic behavior due to the (multiple) sign changes in $R_{\rm H}$.  In contrast, in the tetragonal regime at $x\geq 0.17$, $\cot \theta_{\rm H}$ follows a $T^2$-dependence in the regime where $\rho_{xx}$ is approximately $T$-linear.  We note that the quadratic $T$-dependence of  $\cot \theta_{\rm H}$ implies that the temperature dependence of $R_{\rm H}$ observed for $x = 0.17$ and 0.18 is unlikely to be due to the multi-band effect (See Appendix). Thus, the observed $T^2$-dependence of $\cot \theta_{\rm H}$, when combined with the $T$-linear $\rho_{xx}$, leads us to consider that critical nematic fluctuations are responsible for the deviations from Fermi liquid behavior of the Hall effect.

  \begin{figure*}[t]
 	\includegraphics[width=1.0\linewidth]{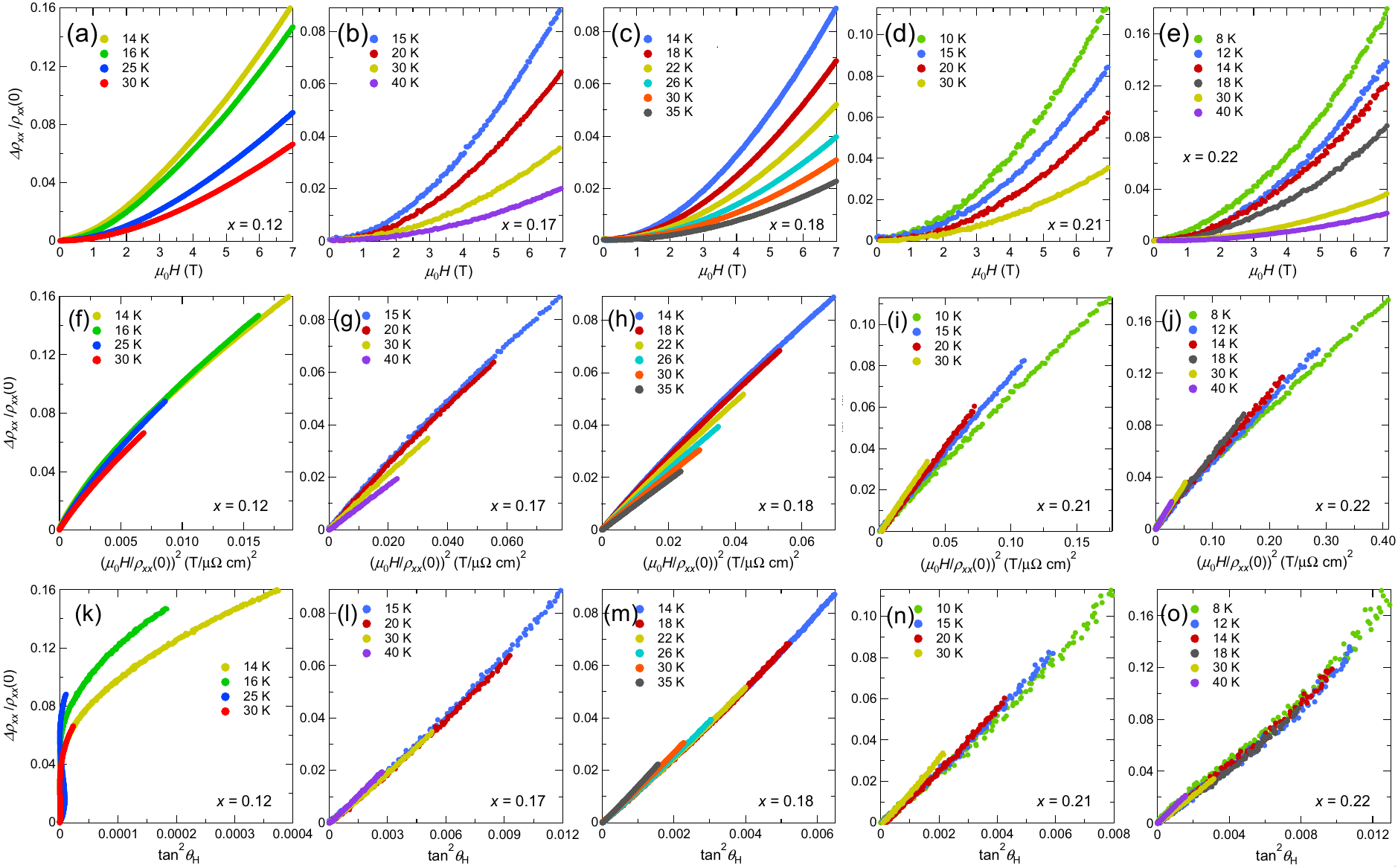}
 	\caption{
 		(a)-(e), Magnetoresistance $\Delta \rho_{xx}(T,H)/\rho_{xx}(T,0)$ of FeSe$_{1-x}$S$_x$ at different doping S substitution levels.  (f)-(j),  $\Delta \rho_{xx}(T,H)/\rho_{xx}(T,0)$  plotted as a function of $[\mu_0H/\rho_{xx}(T,0)]^2$. (k)-(o), $\Delta \rho_{xx}(T,H)/\rho_{xx}(T,0)$ plotted as a function of the square of the tangent of the Hall angle $(\tan \theta_{\rm H})^2$. 
 	}
 	\label{MR}
 \end{figure*}

 \subsection{Magnetoresistance}
The orbital magnetoresistance arises from a bending of the electron trajectory by the Lorentz force. A finite magnetoresistance is a measure of the deviations from a perfectly isotropic Fermi surface, providing unique information on the variation of the electron mean free path around the Fermi surface~\cite{Pippard}. The magnetoresistance of conventional metals is often analyzed in terms of Kohler's rule. According to the semi-classical transport theory based on the Boltzmann equation, Kohler's rule holds when the system has a single carrier and $\tau$ is uniform on the Fermi surface. The field dependence of $\rho_{xx}$ is expressed  by $\omega_c\tau$, where $\omega_c$ is the frequency at which the magnetic field causes the electrons to sweep across the Fermi surface. Since the resistivity in zero field $\rho_{xx}(T,0)$ is inversely proportional to $\tau$, the magnetoresistance of systems with different $\tau$, either due to different impurity levels or temperature, is related to $H$ and $\rho_{xx}(T,0)$.  Thus, the orbital magnetoresistance $\Delta\rho_{xx}(T,H) \equiv \rho_{xx}(T,H)-\rho_{xx}(T,0)$ obeys the relation, 
   
 \begin{equation}
    \frac{\Delta \rho_{xx}(T,H)}{\rho_{xx}(T,0)}=F\left[\frac{H}{\rho_{xx}(T,0)}\right], 
 \end{equation}
 
 \noindent
 where $F(x)$ is a function which depends on details of the Fermi surface.  A violation of Kohler's rule is observed even in conventional metals when the Fermi surface consists of several  different bands and $\tau$  is strongly band dependent.

 Figures\,\ref{MR}(a)-(e) depict the transverse magnetoresistance of FeSe$_{1-x}$S$_x$ for various $x$-values for {\boldmath $H$} $\parallel c$ below 40\,K. To search for Kohler's rule scaling, it is necessary to extract the orbital part of the magnetoresistance.  The orbital part is obtained by subtracting the  longitudinal magnetoresistance measured in the geometry {\boldmath $H$} $\parallel$ {\boldmath $j$} $\parallel ab$  from the transverse one.  As reported in Ref.~\cite{Licciardello19a}, the longitudinal magnetoresistance is negligibly small for $x \geq$ 0.17 [see also Figs.\,\ref{resistivity}(c) and (e)], indicating that the transverse magnetoresistance is essentially dominated by the orbital contribution. 
 
Figures\,\ref{MR}(f)-(j) depict the transverse magnetoresistance below 40\,K plotted as a function of $[\mu_0H/\rho_{xx}(T,0)]^2$ for various $x$-values.  At $x = 0.12$ in the nematic regime, the magnetoresistance collapses onto a single curve below 25\,K, indicating that Kohler's rule is obeyed at low temperatures.   On the other hand, for $x\geq0.17$ in the tetragonal regime, $\Delta\rho_{xx}(T,H)/\rho_{xx}(T,0)$ is not scaled by $\mu_0H/\rho_{xx}(T,0)$, indicating an apparent violation of Kohler's rule.  

In cuprates,  BaFe$_2$(As$_{1-x}$P$_x$)$_2$ and Ce$M$In$_5$ near the AFM QCP, the magnetoresistance  strongly violates Kohler's rule and a new scaling relation,
 \begin{equation}
 	\frac{\Delta\rho_{xx}(T,H)}{\rho_{xx}(T,0)}=p \tan^2 \theta_{\rm H},
 	\label{MKohlereq}
 \end{equation}
where $p$ is a constant, has been shown to hold~\cite{Harris95,Nakajima07,Kasahara10}. This modified Kohler's rule states that the $T$- and $H$- dependencies of $\Delta\rho_{xx}/\rho_{xx}$ collapse onto a single straight line as a function of $\tan^2 \theta_{\rm H}$. Figures\,\ref{MR}(f)-(g)  display $\Delta\rho_{xx}/\rho_{xx}$ plotted as a function of $\tan^2 \theta_{\rm H}$ for various $x$-values.  For $x=0.12$,  $\Delta\rho_{xx}/\rho_{xx}$  is not scaled by $\tan^2 \theta_{\rm H}$. Two remarkable features are found in $\Delta\rho_{xx}/\rho_{xx}$ for $x=0.17$, 0.18 and 0.21 however.  Firstly, $\Delta\rho_{xx}/\rho_{xx}$ is well scaled by $\tan^2 \theta_{\rm H}$.  Secondly, the scaling function is almost perfectly linear in $\tan^2 \theta_{\rm H}$.  These results demonstrate that the low-field magnetoresistance obeys the modified Kohler's rule. Thus, in the regime where deviations from Fermi liquid behavior are observed in the resistivity and Hall effect, the magnetoresistance violates Kohler's rule and can be well described by the modified Kohler's rule. With increasing $x$, Kohler's rule begins to be recovered at low fields as seen in $x=0.22$ [Fig.\,4(j)]. This is consistent with the recovery of a Fermi liquid ground state as manifest in the $T^2$-dependence of $\rho_{xx}$ (the blue region in Fig.\,2).

\section{Discussion}

Several notable attempts have been made to explain these anomalous deviations from Fermi liquid transport, but mostly in systems that display magnetic quantum criticality.   Among others, one may invoke different electron scattering times at different parts of the Fermi surface. However, as the Fermi surface smoothly changes when crossing the nematic QCP~\cite{Hanaguri18,Coldea19}, it is unlikely that the electron scattering time dramatically changes at $x\approx0.17$.   An important effect may arise from vertex corrections to the longitudinal and transverse conductivities due to large AFM fluctuations, which modify the current at the Fermi surface spots connected via the nesting vectors {\boldmath $q$}$_{\rm AF}$ \cite{Stojkovic97,Kontani99, Kontani08}. This effect is pronounced in the vicinity of the AFM QCP, where the critical AFM fluctuations are enhanced.   It has been shown that such vertex corrections give rise to the $T$-linear $\rho_{xx}$, 
an inverse Hall angle that is proportional to $T^2$, and a magnetoresistance that is scaled by $\tan \theta_{\rm H}$. This  scenario requires a large amplitude of {\boldmath $q$}$_{\rm AF}$. In cuprates, iron-pnictides and Ce$M$In$_5$,  AFM spin fluctuations are peaked at {\boldmath $q$}$_{\rm AF}$ connecting certain points on the Fermi surface. In FeSe$_{1-x}$S$_x$, however, the nematic fluctuations are peaked at {\boldmath$q$}$_{\rm nem}\approx 0$ as its ordered state breaks the rotational symmetry while preserving translational symmetry. Thus, it is an open question whether the nematic fluctuations in FeSe$_{1-x}$S$_x$ can give rise to the observed  deviations from Fermi liquid behavior through vertex corrections alone.

Near a nematic QCP, the effective electron-electron interactions are believed to become long ranged, rendering the whole of the Fermi surface `hot', the electronic states incoherent and the transport properties NFL-like~\cite{Mousatov19}. It has been argued, however, that a correct treatment of the electron-phonon coupling in such systems will preserve the integrity of the quasiparticles and recover Fermi liquid behavior~\cite{Paul17, deCarvalho19}. Our results suggest that such considerations are not applicable to FeSe$_{1-x}$S$_x$, possibly due to a diminishing role of electron-phonon coupling in the presence of strongly electron correlations. Indeed, a common set of anomalous charge transport properties observed in  strongly correlated electron systems having essentially different types of critical quantum fluctuations  appears to capture an universal feature of the NFL transport properties near the QCP. Further investigation is required  to clarify this issue.

\section{Conclusion}
In summary, we demonstrate anomalous low-field transport properties in FeSe$_{1-x}$S$_x$ superconductor, which has a QCP of an electronic nematic order that is not accompanied by sizable magnetic fluctuations. The set of anomalous transport properties resemble to those reported in other correlated materials in the vicinity of AFM QCP. The common noticeable features of the anomalous transport behavior observed in correlated electron systems having essentially different types of critical quantum fluctuations capture a universality of NFL behavior near the QCP.

\bigskip

\section*{ACKNOWLEDGMENTS}
We thank  N. Bari\v{s}i\'{c}, E. Berg, A. I. Coldea, E.-G. Moon,  R. Peters and Y. Yamakawa for fruitful discussions. This work is supported by Grants-in-Aid for Scientific Research (KAKENHI) (Nos. JP15H03688, JP15KK0160, JP18H01175, JP18H01177, JP18H05227, JP19H00649, JP20H02600, JP20H05162 and JP20K21139) and on Innovative Areas  ``Topological Material Science" (No. JP15H05852) and ``Quantum Liquid Crystals" (No. JP19H05824) from Japan Society for the Promotion of Science (JSPS), and JST CREST (JPMJCR19T5). We also acknowledge the support by the High Field Magnet Laboratory (HFML) at Radboud University (RU), members of the European Magnetic Field Laboratory (EMFL), and the former Foundation for Fundamental Research on Matter (FOM), which is financially supported by the Netherlands Organisation for Scientific Research (NWO) (Grant No. 16METL01) -- ``Strange Metals". Part of this work was also supported by the European Research Council (ERC) under the European Union's Horizon 2020 research and innovation
programme (Grant agreement No. 835279-Catch-22).

\appendix

\section{Transport coefficients calculated by a two band model}

Here, we demonstrate that the anomalous transport properties 
observed in the vicinity of the nematic QCP in FeSe$_{1-x}$S$_x$ 
cannot be ascribed to  a simple multi-band effect. 
FeSe$_{1-x}$S$_x$ is a compensated metal, 
whose Fermi surface consists of hole and electron pockets.  
By taking this situation into account, we calculate the transport coefficients assuming a two-band model
with circular electron- and hole-pockets.
In the relaxation time approximation, 
the electrical conductivity $\sigma_{xx}$, 
Hall conductivity $\sigma_{xy}$, 
and magnetoconductivity $\Delta \sigma_{xx}$ are respectively given as

\begin{equation}
\sigma_{xx} = ne^2\left(\frac{\tau_h}{m_h} + \frac{\tau_e}{m_e}\right),
\end{equation}
\begin{equation}
\sigma_{xy} = ne^3\left(\frac{\tau_h^2}{m_h^2} - \frac{\tau_e^2}{m_e^2}\right)H,
\end{equation}
\begin{equation}
\Delta\sigma_{xx} = -ne^4\left(\frac{\tau_h^3}{m_h^3} + \frac{\tau_e^3}{m_e^3}\right)H^2,
\end{equation}
where $n$ is the carrier number of each hole- and electron-pocket, 
$\tau_h$ ($\tau_e$) is the scattering time, 
and $m_h$ ($m_e$) is the effective mass of the hole (electron) pocket. 
Then, Hall coefficient, the cotangent of the Hall angle, 
and the magnetoresistance are respectively given as
\begin{equation}
R_H = \frac{\sigma_{xy}}{\sigma_{xx}^2}, 
\end{equation}
\begin{equation}
\cot \theta_H = \frac{\sigma_{xx}}{\sigma_{xy}},
\end{equation}
\begin{equation}
\frac{\Delta\rho_{xx}}{\rho_{xx}(0)} = -\frac{\Delta\sigma_{xx}}{\sigma_{xx}} - \left(\frac{\sigma_{xy}}{\sigma_{xx}}\right)^2.  
\end{equation}

To examine the temperature dependences of these transport coefficients, we assume power-law temperature dependences of the scattering times, $1/\tau_e \propto T^{\eta}$ and $1/\tau_h \propto T^{\zeta}$.  As $R_H$ becomes temperature independent for the exponents $\eta=\zeta$,  we assume $\eta\ne\zeta$. To reproduce the observed $T$-linear resistivity, the relations $\eta < 1$ and $\zeta > 1$ are required. In addition, to reproduce the experimental $T$-dependence of $R_H$, $\tau_h/m_h\sim \tau_e/m_e$ is required at high temperatures ($R_H\sim0$), and $\tau_h/m_h \gg  \tau_e/m_e$ at low temperatures ($R_H>0$).

To satisfy these  constraints, here we set $\eta = 0.7$, $\zeta = 1.3$, $m_e = 2m_0$ and $m_h = m_0$, where $m_0$ is the bare electron mass. Figure\,5(a) shows the $T$-dependences of $1/\tau_e$ and $1/\tau_h$, and Figs.\,5(b)-(e) show the obtained $T$-dependences of $\rho_{xx}$, $R_H$, $\cot \theta_H$ and $\Delta \rho_{xx}/\rho_{xx}(0)$, respectively. As shown in Figs.\,5(b) and (c), the observed $T$-linear resistivity and the low-temperature enhancement of $R_H$ can be reproduced qualitatively using these parameters. However, as shown in Fig.\,5(d) and its inset, $\cot \theta_H$ strongly deviates from a $T^2$-dependence. Moreover, as shown in Fig.\,5(e) and its inset, $\Delta\rho_{xx}/\rho_{xx}(0)$ cannot be scaled by $\tan^2\theta_H$. Thus, these calculations demonstrate that the modified Kohler's rule is strongly violated for the simple two-band model with a $T$-linear resistivity and a low-temperature enhancement of $R_H$.

To summarize, the observed transport properties, $\rho_{xx}=\rho_0 + AT$, $\cot \theta_H = a+bT^2$ and 
$\Delta \rho_{xx}/\rho_{xx}(0)=p\tan^2\theta_H$ cannot be satisfied simultaneously within the relaxation time approximation.   
These results indicate the possible role of vertex corrections in explaining experimental non-Fermi liquid transport coefficients.

\begin{figure}[h]
	\includegraphics[width=0.99\linewidth]{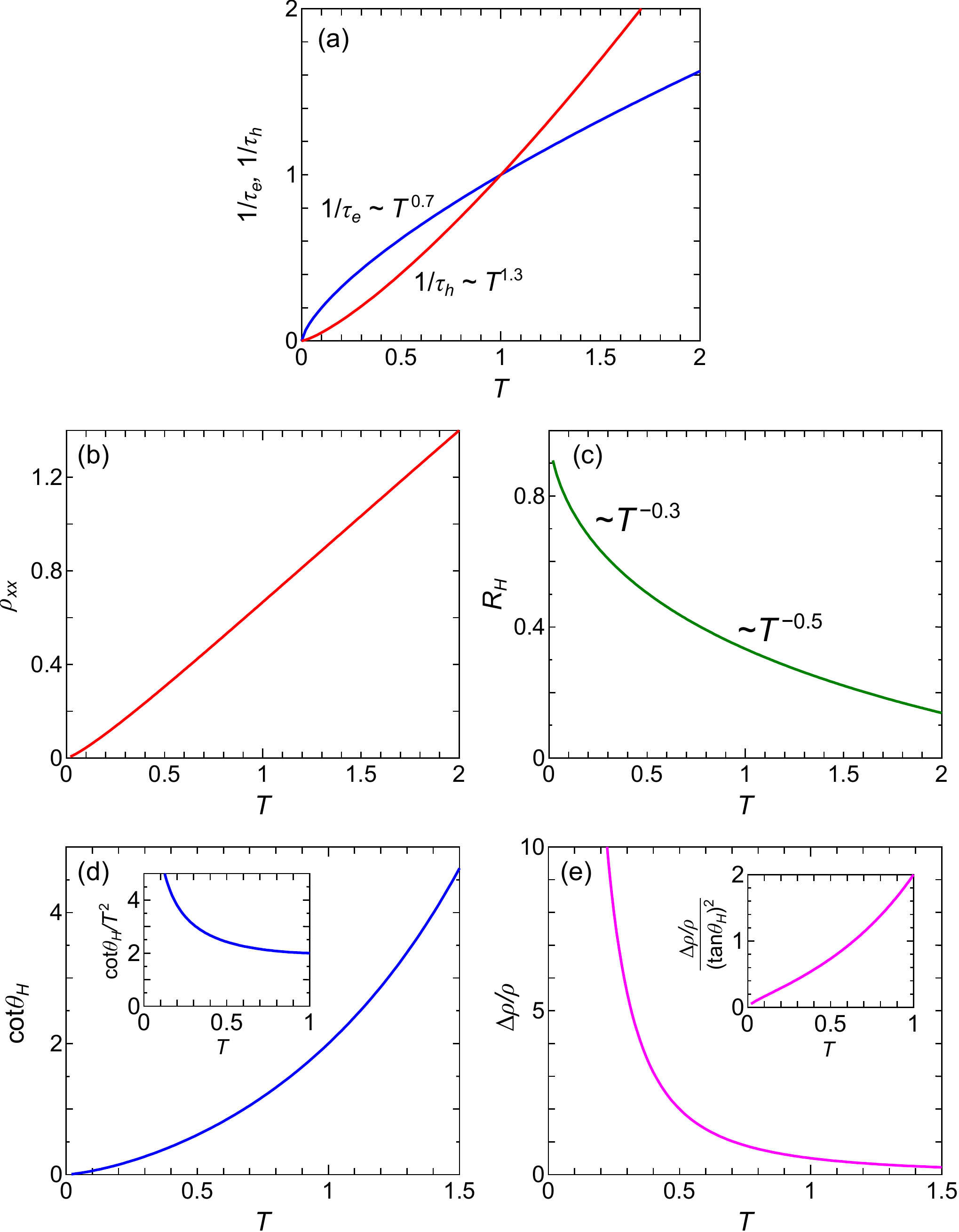}
	\caption{
		Transport coefficients in a two-band model. 
		(a) Temperature dependences of $1/\tau_e$ and $1/\tau_h$, 
		(b) $\rho_{xx}$, (c) $R_H$, (d) $\cot\theta_H$, and (e) $\Delta\rho_{xx}/\rho_{xx}(0)$, calculated by using $\eta = 0.7$, $\zeta= 1.3$, $m_e = 2m_0$ and $m_h = m_0$. 
		The inset in (d) shows a plot of $\cot\theta_H/T^2$ as a function of $T$. The inset in (e) shows a plot of $(\Delta\rho/\rho)/(\tan\theta_H)^2$ as a function of $T$.  Note that $\cot\theta_H$ strongly deviates from a $T^2$-dependence.	Moreover, $(\Delta\rho/\rho)/(\tan\theta_H)^2$ changes with temperature, demonstrating the violation of the modified Kohler's rule in this two-band picture. 
	}
\end{figure}


\bigskip




\begin{thebibliography}{99}
\bibitem{Lohneysen07}
H.\,v. L\"{o}hneysen, A. Rosch, M. Vojta, and P. Wolfle, 
Fermi-liquid instabilities at magnetic quantum phase transitions.
{\rm Rev. Mod. Phys.} {\bf 79}, 1015 (2007). 

\bibitem{Shibauchi14}
T. Shibauchi, A. Carrington,  and Y. Matsuda,  
A quantum critical point lying beneath the superconducting dome in iron pnictides. 
{\rm Annu. Rev. Condens. Matter Phys.} {\bf 5}, 113 (2014).


\bibitem{Martin90}
S. Martin, A. T. Fiory, R. M. Fleming, L. F. Schneemeyer, and J. V. Waszczak, 
Normal-state transport properties of Bi$_{2+x}$Sr$_{2-y}$CuO$_{6+\delta}$ single crystals. 
{\rm Phys. Rev. B} {\bf 41}, 846?849 (1990).

\bibitem{Chien91}
T. R. Chien, Z. Z. Wang, and N.\,P. Ong, 
Effect of Zn impurities on the normal-state Hall angle in single-crystal YBa$_2$Cu$_{3-x}$Zn$_x$O$_{7-\delta}$.
{\rm Phys. Rev. Lett.} {\bf 67}, 2088  (1991).


\bibitem{Harris95}
J. M. Harris, Y. F. Yan, P. Matl, N. P. Ong, P. W. Anderson, T. Kimura, and K. Kitazawa.
Violation of Kohler's Rule in the Normal-State Magnetoresistance of YBa$_2$Cu$_3$O$_{7-\delta}$ and La$_2$Sr$_x$CuO$_4$.
{\rm Phys. Rev. Lett.}{\bf 75}, 1391(1995).

\bibitem{Kasahara10}
S. Kasahara, T. Shibauchi, K. Hashimoto, K. Ikada, S. Tonegawa, R. Okazaki, H. Ikeda, H. Takeya, K. Hirata, T. Terashima, Y. Matsuda, 
Evolution from non-Fermi- to Fermi-liquid transport via isovalent doping in BaFe$_2$(As$_{1-x}$P$_x$)$_2$ superconductors. 
{\rm Phys. Rev. B} {\bf 81}, 184519 (2010).


\bibitem{Nakajima07}
Y. Nakajima, H. Shishido, H. Nakai, T. Shibauchi, K. Behnia, K. Izawa, M. Hedo, Y. Uwatoko, T. Matsumoto, R. Settai, Y. Onuki, H. Kontani, and Y. Matsuda, 
Non-Fermi Liquid Behavior in the Magnetotransport of Ce$M$In$_5$ ($M$: Co and Rh): Striking Similarity between Quasi Two-Dimensional Heavy Fermion and High-$T_c$ Cuprates. 
{\rm J. Phys. Soc. Jpn.} {\bf 76}, 024703 (2007).

\bibitem{Bruin13}
J. A. N. Bruin, H. Sakai, R. S. Perry, and A. P. Mackenzie, 
Similarity of scattering rates in metals showing $T$-linear resistivity.  
{\rm Science} {\bf 339}, 804-807 (2013).

\bibitem{Legros19}
A. Legros, S. Benhabib, W. Tabis, F. Lalibert\'{e}, M. Dion, M. Lizaire, B. Vignolle, D. Vignolles, H. Raffy, Z. Z. Li, P. Auban-Senzier, N. Doiron-Leyraud, P. Fournier, D. Colson, L. Taillefer, and C. Proust, 
Universal $T$-linear resistivity and Planckian dissipation in overdoped cuprates. 
{\rm Nat. Phys.} {\bf 15}, 142-147 (2019).


\bibitem{Licciardello19a}
S. Licciardello, J. Buhot, J. Lu, J. Ayres, S. Kasahara, Y. Matsuda, T. Shibauchi, and N. E. Hussey, 
Electrical resistivity across a nematic quantum critical point. 
{\rm Nature} {\bf 567}, 213-217 (2019). 







\bibitem{Stojkovic97}
B.\,P. Stojkovic and D. Pines, 
Theory of the longitudinal and Hall conductivities of the cuprate superconductors.
{\rm Phys. Rev. B} {\bf 55}, 8576 (1997). 

\bibitem{Kontani99}
H. Kontani, K. Kanki and K. Ueda,
	Hall effect and resistivity in high-$T_c$ superconductors: The conserving approximation. 
	Phys. Rev. B {\bf 59}, 14723 (1999).

\bibitem{Kontani08}
H. Kontani,  
Anomalous transport phenomena in Fermi liquids with strong magnetic fluctuations. 
{\rm Rep. Prog. Phys.} {\bf 71}, 026501 (2008).

\bibitem{Fradkin10}
E. Fradkin, S.\,A. Kivelson, M.\,J. Lawler, J.\,P. Eisenstein, and A.\,P. Mackenzie,  
Nematic Fermi fluids in condensed matter physics. 
{\rm Annu. Rev. Condens. Matter Phys.} {\bf 1}, 153-178 (2010).

\bibitem{Sato17}
Y. Sato, S. Kasahara, H. Murayama, Y. Kasahara, E.-G. Moon, T. Nishizaki, T. Loew, J. Porras, B. Keimer, T. Shibauchi and Y. Matsuda, 
Thermodynamic evidence for a nematic phase transition at the onset of the pseudogap in YBa$_2$Cu$_3$O$_y$. 
{\rm Nat. Phys.} {\bf 13}, 1074-1078 (2017). 

\bibitem{Murayama19}
H. Murayama, Y. Sato, R. Kurihara, S. Kasahara, Y. Mizukami, Y. Kasahara, H. Uchiyama, A. Yamamoto, E.-G. Moon, J. Cai, J. Freyermuth, M. Greven, T. Shibauchi, and Y. Matsuda
Diagonal nematicity in the pseudogap phase of HgBa$_2$CuO$_{4+\delta}$.
{\rm Nat. Commun.} {\bf 10}, 3282 (2019).

\bibitem{Ishida19}
K. Ishida, S. Hosoi, Y. Teramoto, T. Usui, Y. Mizukami, K. Itaka, Y. Matsuda, T. Watanabe, and T. Shibauchi, 
Divergent nematic susceptibility near the pseudogap critical point in a cuprate superconductor. 
J. Phys. Soc. Jpn. {\bf 89}, 064707 (2020). 

\bibitem{Auvray19} 
N. Auvray, B. Loret, S. Benhabib, M. Cazayous, R. D. Zhong, J. Schneeloch, G. D. Gu, A. Forget, D. Colson, I. Paul, A. Sacuto and Y. Gallais,  
Nematic fluctuations in the cuprate superconductor Bi$_2$Sr$_2$CaCu$_2$O$_{8+\delta}$.
{\rm Nat. Commun.} {\bf 10}, 5209 (2019).

\bibitem{Fernandes14}
R. M. Fernandes, A. V. Chubukov, and J.  Schmalian,  
What drives nematic order in iron-based superconductors?
{\rm Nat. Phys.}, {\bf 10}, 97 (2014).

\bibitem{Chu10}
J.-H. Chu, J. G. Analytis, K. D. Greve, P. L. McMahon, Z. Islam, Y. Yamamoto, and I. R. Fisher,
In-plane resistivity anisotropy in an underdoped iron arsenide superconductor. 
{\rm Science} {\bf 329}, 824-826 (2010).

\bibitem{Kasahara12}
S. Kasahara, H. J. Shi, K. Hashimoto, S. Tonegawa, Y. Mizukami, T. Shibauchi, K. Sugimoto, T. Fukuda, T. Terashima, A. H. Nevidomskyy, and Y. Matsuda, 
Electronic nematicity above the structural and superconducting transition in BaFe$_2$(As$_{1-x}$P$_x$)$_2$. 
{\rm Nature} {\bf 486}, 382-385 (2012).

\bibitem{Ronning17}
F. Ronning, T. Helm, K. R. Shirer, M. D. Bachmann, L. Balicas, M. K. Chan, B. J. Ramshaw, R. D. McDonald, F. F. Balakirev, M. Jaime, E. D. Bauer, and P.\,J.\,W.,Moll,  
Electronic in-plane symmetry breaking at field-tuned quantum criticality in CeRhIn$_5$. 
{\rm Nature} {\bf 548}, 313-317 (2017).


\bibitem{Lederer17}
S. Lederer, Y. Schattner, E. Berg, and S.\,A. Kivelson, 
A. Superconductivity and non-Fermi liquid behavior near a nematic quantum critical point. 
{\rm Proc. Natl Acad. Sci. USA}  {\bf 114}, 4905-4910 (2017).

\bibitem{SHM}
T. Shibauchi, T. Hanaguri, and Y. Matsuda,
Exotic Superconducting States in FeSe-based Materials.
arXiv:2005.07315. 


\bibitem{Baek15}
S.-H. Baek, D. V. Efremov, J. M. Ok, J. S. Kim, Jeroen van den Brink, and B. B\"{u}chner, 
Orbital-driven nematicity in FeSe. 
{\rm Nat. Mater.} {\bf 14}, 210-214 (2015).

\bibitem{Bohmer15}
A.\,E. B\"{o}hmer, T. Arai, F. Hardy, T. Hattori, T. Iye, T. Wolf, H.\,v. L\"{o}hneysen, K. Ishida, and C. Meingast, 
Origin of the tetragonal-to-orthorhombic phase transition in FeSe: a combined thermodynamic and NMR study of nematicity. 
{\rm Phys. Rev. Lett.} {\bf 114}, 027001 (2015).


\bibitem{Hosoi16}
S. Hosoi, K. Matsuura, K. Ishida, H. Wang, Y. Mizukami, T. Watashige, S. Kasahara, Y. Matsuda, and T. Shibauchi, 
Nematic quantum critical point without magnetism in FeSe$_{1-x}$S$_x$ superconductors. 
{\rm Proc. Natl Acad. Sci. USA} {\bf 113}, 8139-8143 (2016).


\bibitem{Wiecki17}
P. Wiecki, K. Rana, A. E. B\"{o}hmer, Y. Lee, S. L. Bud'ko, P. C. Canfield, and Y. Furukawa, 
Persistent correlation between superconductivity and antiferromagnetic fluctuations near a nematic quantum critical point in 
Fe$_{1-x}$Se$_x$.
{\rm Phys. Rev. B}  {\bf 98}, 020507(R) (2018).

\bibitem{Coldea19}
A. I. Coldea, S. F. Blake, S. Kasahara, A. A. Haghighirad, M. D. Watson, W. Knafo, E. S. Choi, A. McCollam, P. Reiss, T. Yamashita, M. Bruma, S. Speller, Y. Matsuda, T. Wolf, T. Shibauchi, and A. J. Schofield, 
Evolution of the Fermi surface of the nematic superconductors FeSe$_{1-x}$S$_x$. 
{\rm npj Quant. Mater.} {\bf 4}, 2 (2019).

\bibitem{Hanaguri18}
T. Hanaguri, K. Iwaya, Y. Kohsaka, T. Machida, T. Watashige, S. Kasahara, T. Shibauchi, and Y. Matsuda, 
Two distinct superconducting pairing states divided by the nematic end point in FeSe$_{1-x}$S$_x$. 
{\rm Sci. Adv.} {\bf 4}, eaar6419 (2018).

\bibitem{Sato18}
Y. Sato, S. Kasahara, T. Taniguchi, X.Z. Xing, Y. Kasahara, Y. Tokiwa, Y. Yamakawa, H. Kontani, T. Shibauchi, and Y. Matsuda, 
Abrupt change of the superconducting gap structure at the nematic critical point in FeSe$_{1-x}$S$_x$. 
{\rm Proc. Natl. Acad. Sci. USA} {\bf 115}, 1227-1231 (2018).


\bibitem{Bristow19}
M. Bristow, P. Reiss, A.\,A. Haghighirad, Z. Zajicek, S.\,J. Singh, T. Wolf, D. Graf, W. Knafo, A. McCollam, and A.\,I. Coldea, 
Anomalous high-magnetic field electronic state of the nematic superconductors FeSe$_{1-x}$S$_x$. 
{\rm Phys. Rev. Research} {\bf 2}, 013309 (2020). 

\bibitem{Urata16}
T. Urata, Y. Tanabe, K.\,K. Huynh, H. Oguro, K. Watanabe, and K. Tanigaki, 
Non-Fermi liquid behavior of electrical resistivity close to the nematic critical point in Fe$_{1-x}$Co$_x$Se and FeSe$_{1-y}$S$_y$. 
arXiv:1608.01044. 


\bibitem{Licciardello19b}
S. Licciardello, N. Maksimovic, J. Ayres, J. Buhot, M. \v{C}ulo, B. Bryant, S. Kasahara, Y. Matsuda, T. Shibauchi, V. Nagarajan, J. G. Analytis, and N. E. Hussey, 
Coexistence of orbital and quantum critical magnetoresistance in FeSe$_{1-x}$S$_x$. 
{\rm Phys. Rev. Research} {\bf 1}, 023011 (2019). 



\bibitem{Berg19} 
E. Berg, S. Lederer, Y. Schattner, and S. Trebst.  
Monte Carlo Studies of Quantum Critical Metals. 
{\rm Annu. Rev. Condens. Matter Phys.} {\bf 10}, 63-84 (2019).


\bibitem{Paul17}
I. Paul and M. Garst,
Lattice Effects on Nematic Quantum Criticality in Metals. 
{\rm Phys. Rev. Lett.} {\bf 118}, 227601 (2017). 

\bibitem{deCarvalho19}
V. S. de Carvalho and R. M. Fernandes, 
Resistivity near a nematic quantum critical point: Impact of acoustic phonons. 
{\rm Phys. Rev. B} {\bf 100}, 115103 (2019). 







\bibitem{Kasahara14}
S. Kasahara, T. Watashige, T. Hanaguri, Y. Kohsaka, T. Yamashita, Y. Shimoyama, Y. Mizukami, R. Endo, H. Ikeda, K. Aoyama, T. Terashima, S. Uji, T. Wolf, H.\,v. L\"{o}hneysen, T. Shibauchi, and Y. Matsuda, 
Field-induced superconducting phase of FeSe in the BCS-BEC cross-over.
{\rm Proc. Natl Acad. Sci. USA} {\bf 111}, 16309-16313 (2014). 

\bibitem{Yi19}
M. Yi, H. Pfau, Y. Zhang, Y. He, H. Wu, T. Chen, Z.\,R. Ye, M. Hashimoto, R. Yu, Q. Si, D.-H. Lee, Pengcheng Dai, Z.-X. Shen, D.\,H. Lu, and R.\,J. Birgeneau, 
The Nematic Energy Scale and the Missing Electron Pocket in FeSe.
{\rm Phys. Rev. X} {\bf 9}, 041049 (2019).

\bibitem{Pippard}
A. Pippard, Magnetoresistance in Metals, Cambridge Studies in Low Temperature Physics (Cambridge University Press, 1989).

\bibitem{Mousatov19}
C.\, H. Mousatov, E. Berg, and S.\,A. Hartnoll, 
Theory of the strange metal Sr$_3$Ru$_2$O$_7$. 
{\rm Proc. Natl Acad. Sci. USA} {\bf 117}, 2852-2857 (2020). 

\end{thebibliography}
\end{document}